\documentstyle[12pt]{article}
\title{Reduction of the spin-orbit potential in light drip-line nuclei}
\author{G.A. Lalazissis, D. Vretenar\footnotemark[1],
W. P\" oschl and P. Ring \\
Physik-Department der Technischen Universit\"at M\"unchen, \\
D-85747 Garching, Germany}
\begin{document}
\maketitle
\footnotetext[1]{Alexander von Humboldt Fellow,
on leave of absence from University of Zagreb, Croatia}
\begin{abstract}
The isospin dependence of the spin-orbit interaction
in light neutron rich nuclei is investigated in 
the framework of relativistic mean field theory.
The magnitude of the spin-orbit 
potential is considerably reduced in drip line nuclei,
resulting in smaller energy splittings between 
spin-orbit partners. The effect does not depend
on the parametrization of the effective Lagrangian.
The results are compared with corresponding calculations
in the non-relativistic Skyrme model.
\end{abstract}

The spin-orbit interaction plays a central role in the physics of
nuclear structure. It is rooted in the basis of the nuclear 
shell model, where its inclusion is essential in order to 
reproduce the experimentally established magic numbers.
In non-relativistic models based on the mean field approximation,
the spin-orbit potential is included in a phenomenological way. 
Of course such an ansatz introduces an additional parameter,
the strength of the spin-orbit interaction. The value of this
parameter is usually adjusted to the experimental 
spin-orbit splittings in spherical nuclei, for example $^{16}$O.
On the other hand, in the relativistic framework
the nucleons are described as Dirac spinors. This means that in the 
relativistic description of the nuclear many-body problem, 
the spin-orbit interaction arises naturally from the
Dirac-Lorenz structure of the effective Lagrangian. No
additional strength parameter is necessary, and 
relativistic models reproduce the empirical
spin-orbit splittings. 

Many  properties of nuclei along the line of beta stability have
been successfully described in the framework of models 
based on the mean-field approximation. Conventional non-relativistic 
models that include
density dependent interactions with finite range (Gogny) \cite{DG.80}, or
zero-range (Skyrme) forces \cite{FQ.78}, have been extensively 
used to describe the structure of stable nuclei. More recently, it has
been shown that models based on the relativistic 
mean-field theory~\cite{SW.86,SW.97} provide an elegant and economical 
framework, in which properties of nuclear matter and finite nuclei,
as well as the dynamics of heavy-ion collisions, can be 
calculated (for a recent review see~\cite{Rin.96}).
In comparison with conventional non-relativistic 
approaches, relativistic models explicitly
include mesonic degrees of freedom and describe the nucleons 
as Dirac particles. 
Non-relativistic models and the relativistic mean-field theory
predict very similar results for many properties of beta stable nuclei.
However, cases have been found where the non-relativistic 
description of nuclear structure fails. An example is the 
anomalous kink in the isotope shifts
of Pb nuclei \cite{SLR.93}. This phenomenon could not be
explained neither by the Skyrme model, nor by the Gogny approach.
Nevertheless, it is reproduced very naturally in relativistic 
mean-field calculations.
A more careful analysis~\cite{lala94} has shown that the origin 
of this discrepancy is the isospin dependence of the spin-orbit term.
With a spin-orbit term modified in such a way that it is
similar to that derived in the relativistic mean-field model,
the Skyrme model produces comparable results  
for the isotope shifts~\cite{SLK.94}.
Recently, in Ref~\cite{RF.95} another
modification in the spin-orbit term of the energy functional of 
the Skyrme model has been proposed. An additional parameter was 
introduced, which is adjusted to
reproduce the kink in the isotope shifts. 
Compared to conventional Skyrme forces, this approach produces
a very different isospin dependence of the spin-orbit potential.

Experiments with radioctive nuclear beams provide the 
opportunity to study nuclei with large neutron excess. Neutron drip lines
of relatively light nuclei have become accesible, and the investigation 
of properties of such exotic objects is becoming one of
the most exciting challenges in the physics of nuclear structure.
Because of their relevance to the r-process in nucleosynthesis,
nuclei close to the neutron drip line are also very important in nuclear
astrophysics. Knowledge of their stucture and properties would 
help the determination of astrophysical conditions for the formation of
neutron rich stable isotopes~\cite{Kra.88}. For drip line isotopes, 
nuclear shell effects become very important and the spin-orbit term plays
an essential role. Very different scenarios for the isospin dependence
of the spin-orbit interaction have been suggested in the
Skyrme model~\cite{Doba.94}, and in the relativistic
mean-field theory~\cite{SLH.94}. In the present work we investigate
the behaviour of the spin-orbit potential in light neutron rich nuclei.
The description of drip line nuclei is 
complicated by the closeness of the Fermi level to 
continuum states. Pairing correlations, and the 
coupling between bound states and positive energy particle continuum,
are described in the Relativistic Hartree-Bogoliubov (RHB) model
in coordinate space.

In the Hartree approximation for the self-consistent
mean field, the RHB equations read
\begin{eqnarray}
\left( \matrix{ \hat h_D -m- \lambda & \hat\Delta \cr
                -\hat\Delta^* & -\hat h_D + m +\lambda
                         } \right) \left( \matrix{ U_k \cr V_k } \right) =
E_k\left( \matrix{ U_k \cr V_k } \right).
\end{eqnarray}
where $\hat h_D$ is the single-nucleon Dirac hamiltonian,
and $m$ is the nucleon mass.
$U_k$ and $V_k$ are  quasi-particle Dirac spinors, and $E_k$ denotes
the quasi-particle energies. The RHB equations are non-linear 
integro-differential equations. They have to be solved self-consistently,
with potentials determined in the mean-field approximation from 
solutions of Klein-Gordon equations for mesons. In the particle-particle
($pp$) channel the pairing interaction is approximated by a 
two-body finite range Gogny interaction
\begin{equation}
V^{pp}(1,2)~=~\sum_{1,2}
e^{-( {\bf r}_1- {\bf r}_2
/ {\mu_i} )^2}\,
(W_i~+~B_i P^\sigma 
-H_i P^\tau -
M_i P^\sigma P^\tau),
\end{equation}
with parameters $\mu_i$, $W_i$, $B_i$, $H_i$ and $M_i$ $(i=1,2)$.

In the present investigation we consider 
the isotopic chains of the even-even Ne and Mg nuclei. 
Systematic RHB calculations of ground state properties
have been performed, and the
results reveal many interesting features. In this report 
we focus on
the behaviour of the spin-orbit term. In the relativistic mean-field
approximation, the spin-orbit potential originates from the addition
of two large fields: the field of the vector mesons (short
range repulsion), and the scalar field of the sigma meson (intermediate
attraction). In the first order approximation, and assuming spherical 
symmetry, the spin orbit term can be written as
\begin{equation}
\label{so1}
V_{s.o.} = {1 \over r} {\partial \over \partial r} V_{ls}(r), 
\end{equation} 
where $V_{ls}$ is the spin-orbit potential~\cite{Rin.96,Koepf.91}
\begin{equation}
\label{so2}
V_{ls} = {m \over m_{eff}} (V-S).
\end{equation}
V and S denote the repulsive vector and the 
attractive scalar potentials, respectively.
$m_{eff}$ is the effective mass
\begin{equation}
\label{so3}
m_{eff} = m - {1 \over 2} (V-S).
\end{equation}

In the following we present results for the even-even
Ne and Mg isotopes.  For the mean-field
Lagrangian the NL3~\cite{LKR.96} parametrization has been
used, and the parameter set D1S~\cite{BGG.84} for
the finite range pairing interaction.
Using the vector and scalar potentials from the self-consistent
ground-state solutions, we have computed 
from~(\ref{so1}) - (\ref{so3}) the corresponding spin-orbit terms for Ne and Mg chains.
They are displayed in the upper panels of Fig. 1, as function
of the radial distance from the center of the nucleus. The magnitude 
of the spin-orbit term $V_{s.o.}$ decreases as we add more neutrons, i.e.
more units of isospin. The reduction for nuclei close 
to the neutron drip is $\approx 40\%$ in the surface region,
as compared to values which 
corespond to beta stable nuclei.
This implies a significant weakening of the spin-orbit interaction. 
The minimum of $V_{s.o.}$ is also shifted outwards, and this  
indicates the large spatial extension of the scalar and vector densities,
which become very diffused on the surface. The main contribution 
to the densities in this region comes from the outermost neutron orbitals, 
whose wave functions are extened in space. The same effect can be 
observed in Fig. 2, where the $rms$ radii  
are plotted as function of neutron number for Ne (a), and Mg (b) isotopes.
We display neutron and proton $rms$ radii, and the N$^{1/3}$ curves.
The dashed curves  are normalized so that 
they coincide with neutron $rms$ radii for $^{20}$Ne and $^{24}$Mg, 
respectively.  Close to the neutron drip a sharp increase
of neutron radii is observed, as compared to the mean-field 
curves N$^{1/3}$. This sudden increase 
indicates a possible formation of multi-neutron halo~\cite{Pos.97}. 

In Fig. 3 we display the spin-orbit splittings of the neutron levels
\begin{equation}
 E_{ls} = E_{n,l,j=l-1/2} - E_{n,l,j=l+1/2}, 
\end{equation}
for Ne and Mg isotopes as function of the neutron number.
The neutron spin-orbit splittings decrease with neutron number.
This is consistent with the gradual weakening of the 
spin-orbit term that is shown in Fig. 1. However, the result 
is at variance with non-relativistic mean-field studies that have
used Skyrme forces~\cite{Doba.94}. In the Skyrme model, 
the spin-orbit term included in the self-consistent 
mean field is of the form
\begin{equation}
V^{ls}_{\tau} =  {\bf W}_{\tau}({\bf r})(\bf p \times {\boldmath \sigma}),
\end{equation}
with 
\begin{equation} 
 {\bf W}_{\tau}({\bf r}) = W_{1}{\boldmath \nabla}\rho_{\tau} + 
W_{2}{\boldmath \nabla}\rho_{\tau\prime \ne \tau}
\end{equation}  
where $\rho_\tau$ is the density for neutrons or protons
($\tau=n$ or $p$) and $\tau^\prime$ is the opposite
isospin. $W_{1}$ and $W_{2}$ are parameters.
In the non-relativistic reduction of the Dirac equation
the resulting spin orbit term takes the same form. However, 
the corresponding quantities $W_{1}$ and $W_{2}$ are 
coordinate dependent, and also introduce an 
isospin dependence which has its origin in the 
coupling constant $g_{\rho}$ of the $\rho$-meson~\cite{lala94}
\begin{eqnarray}
W{_1}(r)&=&\frac{1}{4m^2m^{*2}}(C^2_\sigma+C^2_\omega+C^2_\rho)\\
\nonumber
W{_2}(r)&=&\frac{1}{4m^2m^{*2}}(C^2_\sigma+C^2_\omega-C^2_\rho).
\end{eqnarray} 
with $C_i^2=(mg_i/m_i)^2$ for $i=\sigma,\omega,\rho$. 
Non linear self-interaction terms for the sigma meson are included in the
derivation of $W_{1}$ and $W_{2}$.

With the self-consistent proton and neutron densities of the drip line
nucleus $^{40}$Ne, we have computed the neutron spin-orbit terms 
both in the Skyrme model and in the relativistic mean-field model.
Spherical symmetry is assumed, and for simplicity, pairing 
has been neglected in this illustrative calculations. 
The effective force Skp \cite{Doba.84} 
has been used in Skyrme calculations, and the NL3 parameter set for
the effective relativistic Lagrangian.
The results are summarized in Fig. 4. In Figs. 4a,b,d, and e, the neutron
and proton densities, and the corresponding gradients are shown.
The spin-orbit term calculated with the Skyrme force Skp is displayed in
Fig. 4g (solid line), and the one resulting from relativistic calculations
in Fig. 4i. The minimum in the surface region is for the Skyrme force Skp
$\approx 40\%$ deeper than in the relativistic calculations.
In order to have a more complete comparison, in Fig. 4f we also
display results of the relativistic model with 
the NL1 \cite{Rei.89,RRM.86} (dotted line), and NL-SH \cite{SNR.93}
(dashed line) effective forces.
They turn out to be quite similar to those obtained with NL3. The small 
radial shift between the two minima is caused by 
the different isospin properties of
the two parameter sets. The key point, however, is that in all relativistic 
calculations the spin-orbit term is much weaker compared to 
results of the Skyrme model. This of course explains the difference
in the calculated spin-obit splittings. The spin-orbit term in 
the Skyrme model does not decrease in magnitude even for
extremely neutron rich nuclei. This results in large 
spin-obit splittings and quenching of the shell effects.

The Skyrme model uses in principle an
isospin independent two-body spin-orbit force. The exchange 
term to this force, however, induces a strong
isospin dependence for the spin-orbit term in the
self-consistent mean field. The resulting 
ratio for the W$_{i}$ parameters is: W$_{1}$/W$_{2}$=2.
On the other hand, in the relativistic approach
the spin-orbit term is to a large
extent a pure single-particle effect, as can be shown by a 
non-relativistic reduction of the Dirac equation. This
is also the case in  relativistic Hartree-Fock calculations.   
Therefore in the relativistic description there is no contribution 
from a two-body spin-orbit interaction, and consequently no exchange term.
The isospin dependence of the parameters W$_{1}$ and W$_{2}$ in the  
relativistic theory comes from the $\rho$-meson. The contribution 
of the $\rho$ mean-field increases with neutron number. This 
effect is illustrated in the lower panel in Fig. 1,
where we plot the self-consistent $\rho$-meson potential for 
Ne and Mg isotopes. However, its contribution 
is much smaller than that of the $\sigma$ and $\omega$ 
fields, and therefore the 
isospin dependence of W$_{1}$ and W$_{2}$ should be rather weak. 
In Figure 4c we display the relativistic W$_{1}$ and
W$_{2}$ calculated with the effective force NL3. The 
radial dependence comes from the effective mass m$^*$. 
The two curves are very similar and in the
surface region, where the spin-orbit term is peaked,
their ratio is approximately W$_{1}$/W$_{2}$ $\approx$ 1.1. 
If we now use the same ratio in the Skyrme model, the resulting
spin-orbit term for the Skp force is displayed 
in Fig. 4g (dotted line). A significant reduction of
the spin-orbit term is observed, and the results are comparable 
to those obtained in the relativistic calculations. We have
also computed the spin-orbit term for the effective interaction
SkM*~\cite{BQB.82} and the two values of the ratio W$_{1}$/W$_{2}$. The results displayed in Fig. 4h show the same reduction in the 
surface region.

In conclusion, we have shown that, in the framework of 
relativistic mean field theory, the magnitude of the 
spin-orbit potential is considerably reduced in light drip line 
nuclei. With the increase of the neutron number, the effective 
one-body spin-orbit interaction becomes weaker. This result
in a reduction of the energy splittings between spin-orbit partners.
The reduction of the spin-orbit potential is especially pronounced in the 
surface region, and does not depend on a particular parameter set
used for the effective Lagrangian. These results are at variance with
those calculated with the non-relativistic Skyrme model. It has 
been shown that the differences have their origin in the
isospin dependence of the spin-orbit terms in the two  models.
If the spin-orbit term of the Skyrme model is modified in such a way 
that it does not depend so strongly on the isospin,
the reduction of the spin-orbit potential is comparable to 
that observed in relativistic mean-field calculations.

This work has been supported in part by the
Bundesministerium f\"ur Bildung und Forschung under
contract 06~TM~875.

\newpage
{\bf Figure Captions}

\begin{itemize}
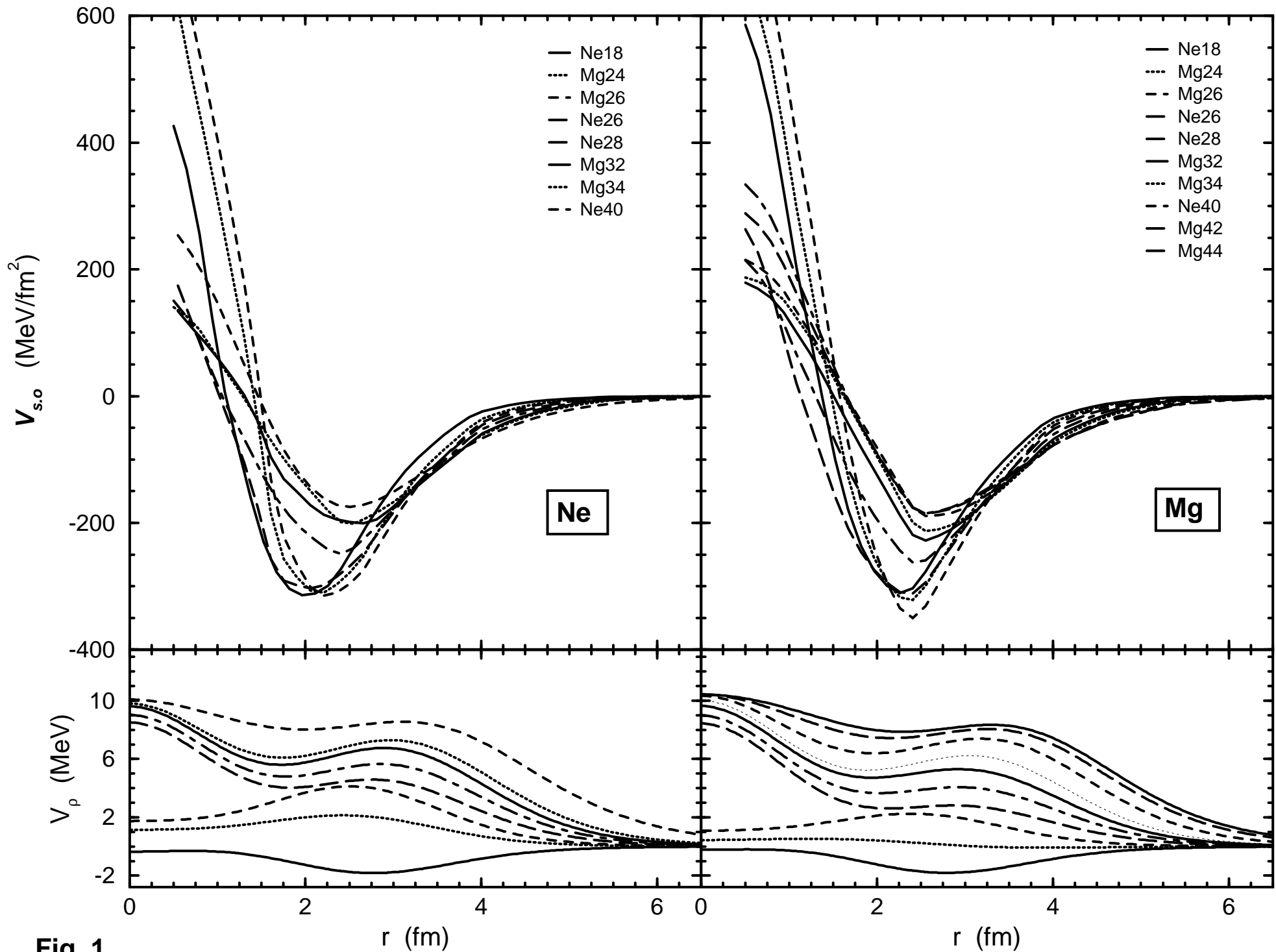
\item{\bf Fig. 1} Radial dependence of the spin-orbit potential
in self-consistent solutions for the ground-states of 
Ne (a), and Mg (b) isotopes. In the lower panel
the self-consistent $\rho$-meson potentials are displayed. 
The NL3 parametrization has been used for the mean-field
Lagrangian, and the parameter set D1S for the pairing interaction.
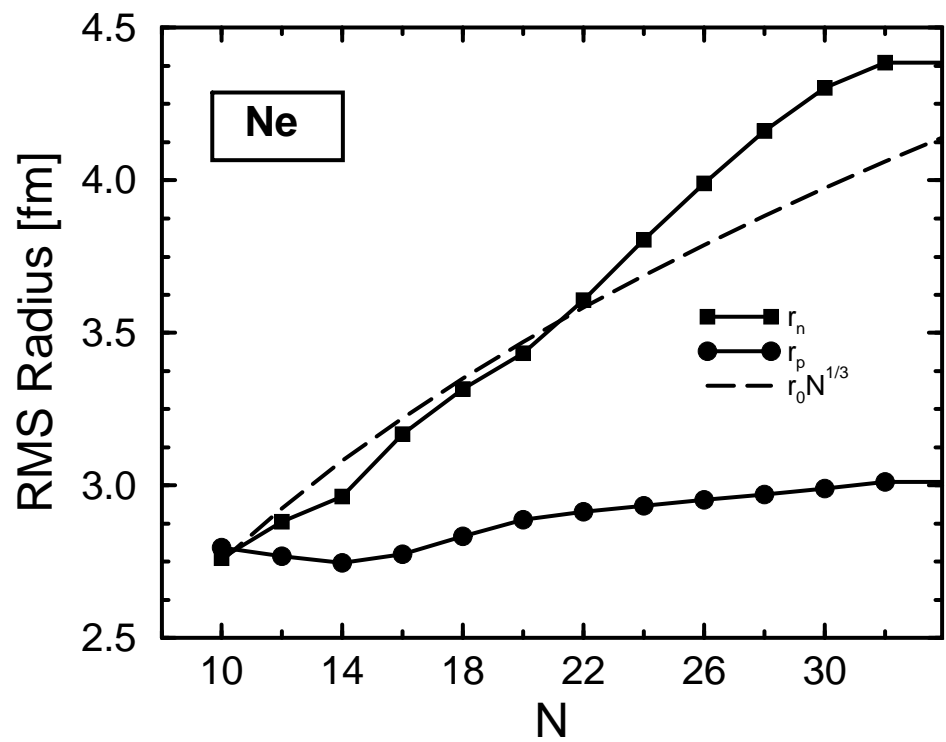
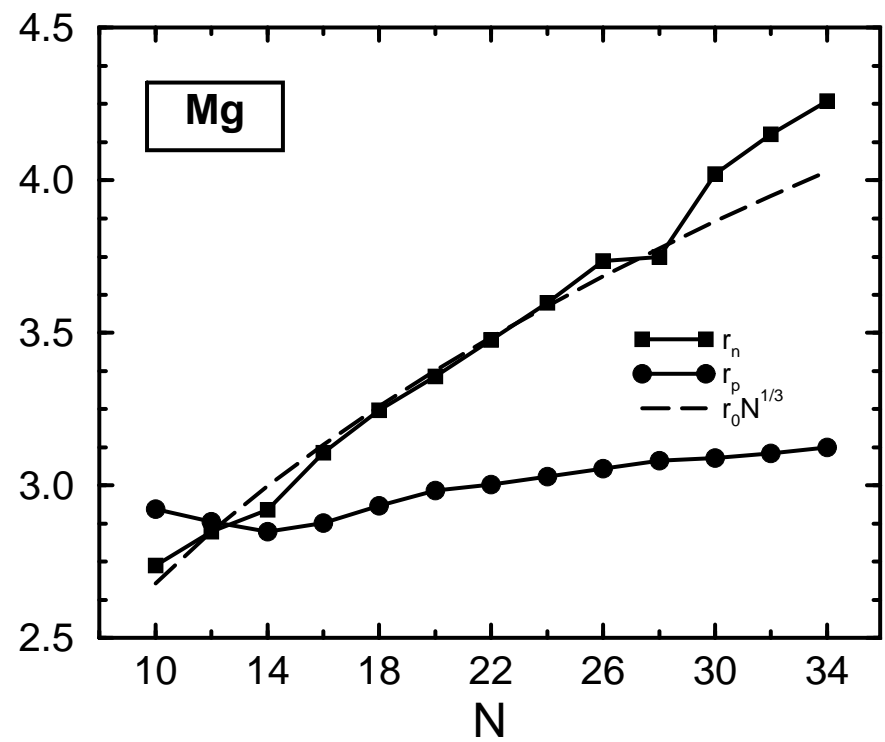
\item{\bf Fig. 2}  Calculated $rms$ radii for Ne (a), and Mg (b) isotopes
as functions of neutron number.
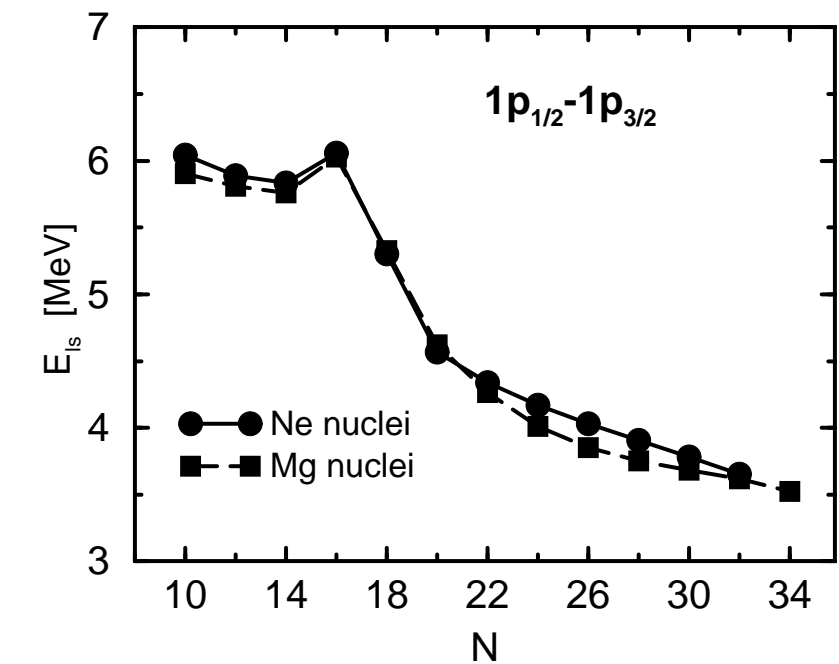
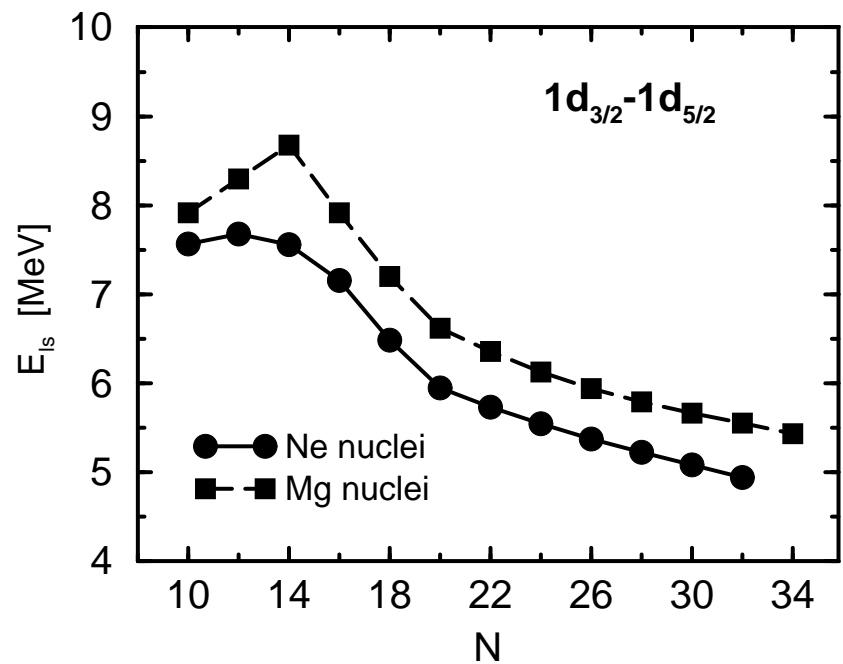
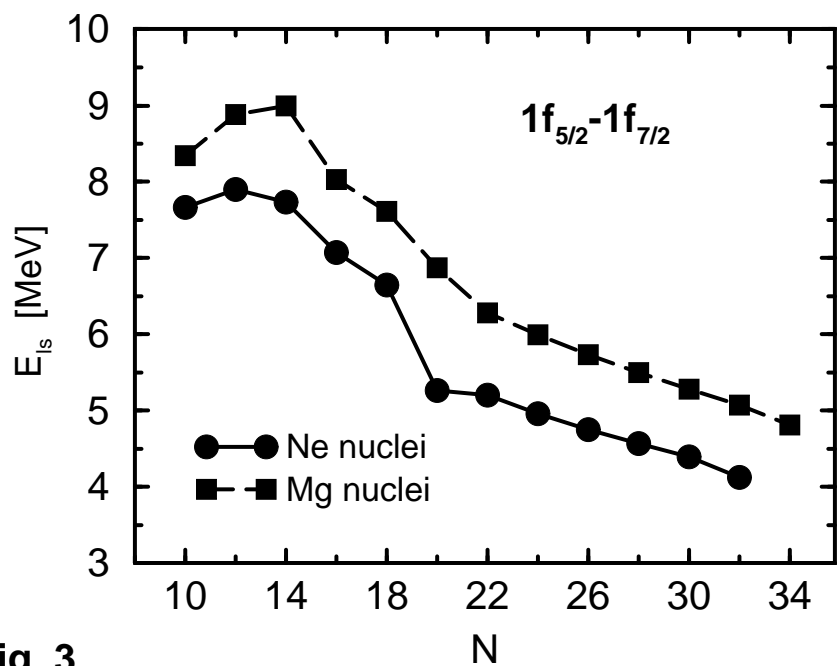
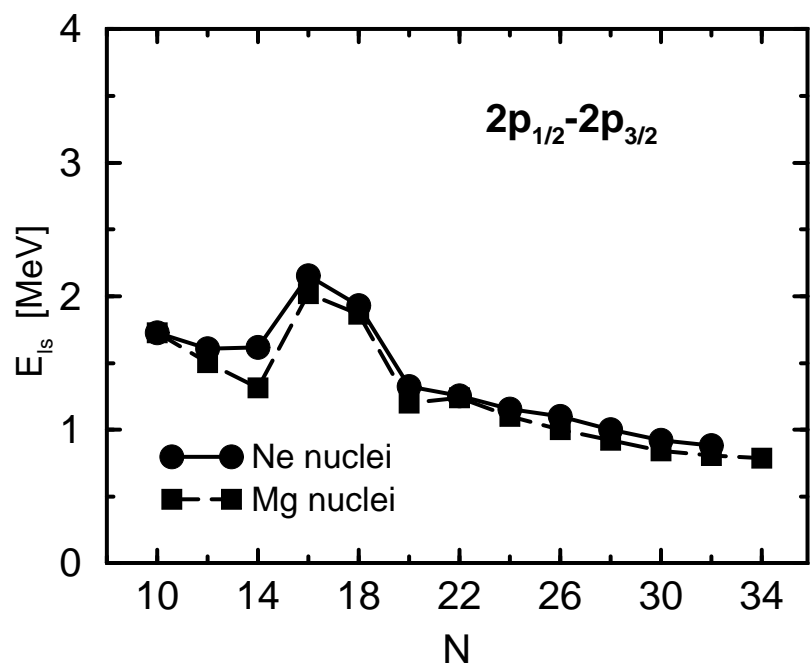
\item{\bf Fig. 3} Energy splittings between spin-orbit partners, 
for neutron levels in Ne and Mg nuclei, as functions of neutron number.
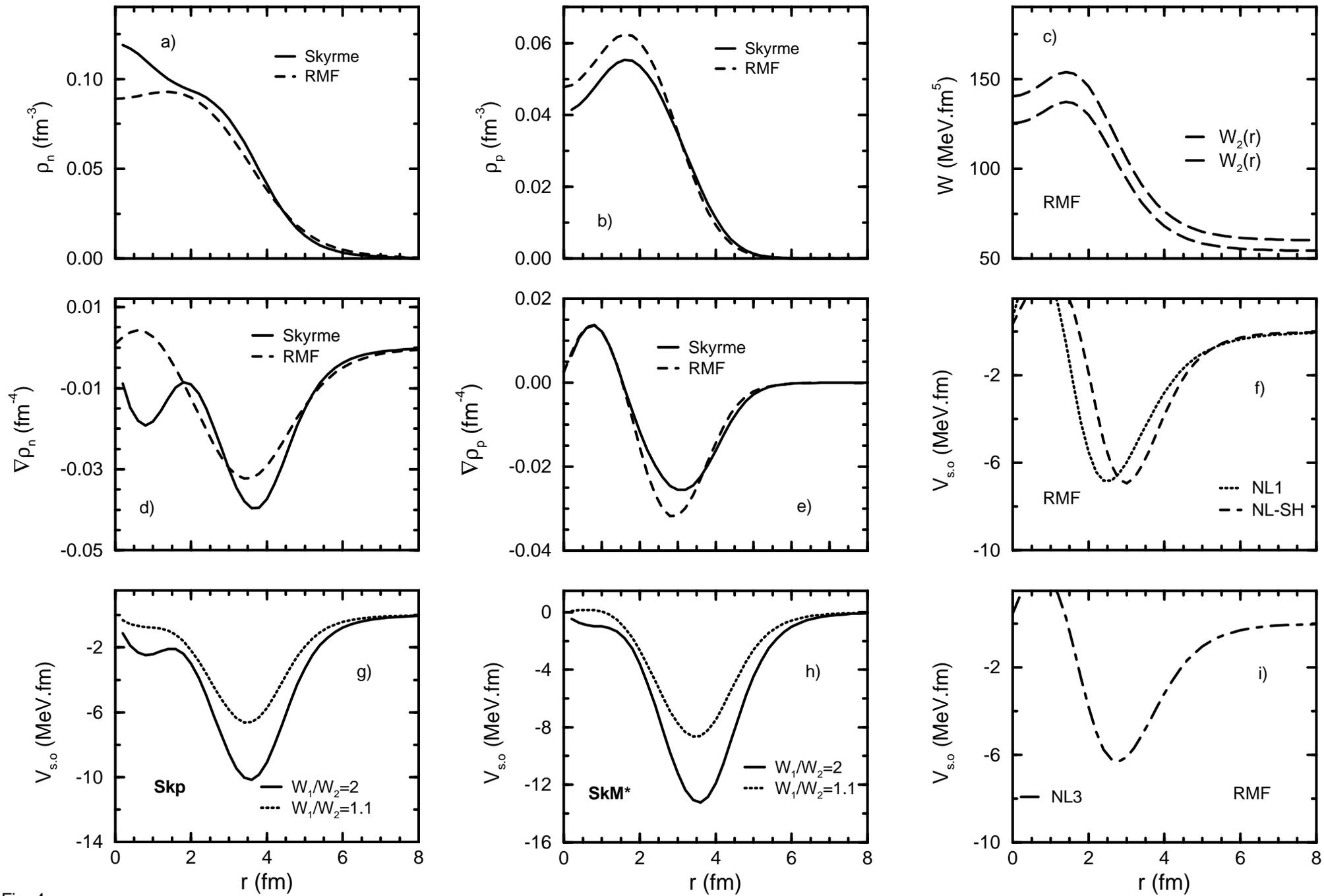
\item{\bf Fig. 4} Self-consistent proton and neutron densities, 
the corresponding gradients, and the spin-orbit terms for 
$^{40}$Ne. Results are displayed for the relativistic 
mean-field model and non-relativistic Skyrme calculations.
For description, see text.
\end{itemize}
\end{document}